\documentstyle[fleqn,run2col,epsfig]{article}
\newcommand{\AmS}{{\protect\the\textfont2
  A\kern-.1667em\lower.5ex\hbox{M}\kern-.125emS}}

\title{
Variable number schemes for heavy quark electroproduction
}

\author{J. Smith\address{C.N. Yang Institute for \\
        Theoretical Physics, \\
        SUNY at Stony Brook , \\
        Stony Brook, NY 11794-3840}%
        \thanks{Work supported in part by the NSF grant PHY-9722101}}

\begin{document}
\begin{abstract}
We compare variable number schemes for heavy quark electroproduction.
\end{abstract}

\maketitle


Heavy quark production has been a major topic of investigation
at hadron-hadron, electron-proton and electron-positron
colliders. Here a review is given of some topics which are of interest
primarily for electron-proton colliders. We concentrate on this
reaction because a theoretical treatment can be based on the 
operator product expansion, and also because data are available 
for deep-inelastic charm production at HERA. How all this relates to 
Fermilab experiments will be discussed at the end. 

In QCD perturbation theory one needs to introduce a renormalization 
scale and a mass factorization scale to perform calculations. 
We choose both equal to $\mu^2$, which will be a function 
of $Q^2$ and the square of the mass of the charm quark $m^2$.
At small $\mu^2$, where kinematic effects due to quark masses are important, 
the best way to describe charm quark production is via heavy quark 
pair production where parton densities are used for the initial state 
light mass u,d,s quarks and the gluon and $m$ only appears in 
the heavy quark coefficient functions (or partonic cross sections) like
$H_{i,g}^{\rm S,(2)}(z,Q^2,m^2,\mu^2)$, etc., \cite{lrsn}. Here the 
superscripts refer to their flavor decomposition and the order in 
perturbation theory, while the subscripts refer to the projection
$i=2,L$ and the partonic initial state. The arguments refer 
to the partonic Bjorken variable $z = Q^2/(s+Q^2)$ 
and to the fact that these functions depend upon invariants and scales. 
The renormalization necessary to calculate
these NLO expressions follows the CWZ method \cite{cwz}. 
The symbol $H$ refers to those coefficient functions which are 
derived from Feynman diagrams where the virtual
photon couples to a heavy quark line. 
Analytic expressions for these functions are not known. 
Numerical fits are available in \cite{rsn}. Asymptotic 
expressions for $Q^2 \gg m^2$ are available in \cite{bmsmn} which contain 
terms like $\ln^2(Q^2/m^2)$ and $\ln(Q^2/m^2)\ln(Q^2/\mu^2)$ 
multipled by functions of $z$ and demonstrate that these functions are 
singular in the limit that $m\rightarrow 0$.
There are other heavy quark coefficient functions such as
$L_{i,q}^{\rm NS,(2)}(z,Q^2,m^2)$, which arise from tree diagrams where the 
virtual photon attaches to the an initial state light quark line, so the
heavy-quark is pair produced via virtual gluons. Analytic expressions
for these functions are known for all $z$, $Q^2$ and $m^2$, which, in the limit
$Q^2 \gg m^2$ contain powers of $\ln(Q^2/m^2)$ multiplied by functions
of $z$. 
The three-flavor light mass $\overline{\rm MS}$ parton densities 
can be defined in terms of matrix elements of operators and are now 
available in parton density sets. This is a fixed order perturbation 
theory (FOPT) description of heavy quark production with three-flavour 
parton densities.  Notice that due to the work in \cite{lrsn} the 
perturbation series is now known up to second order. In regions of 
moderate scales and invariants this NLO description is well defined 
and can be combined with a fragmentation function to predict exclusive 
distributions \cite{hs} for the outgoing charm meson, 
the anti-charm meson and the additional parton.
This NLO massive charm approach agrees well with the recent
$D$-meson inclusive data in \cite{ZEUS} and \cite{H1}. 
Let us call the charm quark structure functions in this NLO description
$F^{\rm EXACT}_{i,c}(x,Q^2,m^2,n_f=3)$, where $i=2,L$. 

A different description, which should be more 
appropriate for large scales, where terms in $m^2$ are 
negligible, is to represent charm production by a parton density 
$f_c(x,\mu^2)$, with a boundary condition that the density vanishes
at small values of $\mu^2$. Although at first sight these approaches 
are completely different they are actually intimately related. 
It was shown in \cite{bmsn1} that the large terms in $\ln(Q^2/m^2)$ 
which arise when $Q^2 \gg m^2$, can be resummed in all orders in perturbation
theory. In this reference all the two-loop corrections to the
matrix elements of massive quark and massless gluon operators
in the operator product expansion were calculated. These contain
the same type of logarithms mentioned above multiplied by functions
of $z$, (which is the last Feynman integration parameter).
After operator renormalization and suitable reorganization of convolutions
of the operator matrix elements (OME's) and the coefficient functions
the expressions for the infrared-safe charm quark 
structure functions $F_{i,c}(x,Q^2,m^2,\Delta)$, 
become, after resummation, convolutions of light mass
four-flavor parton coefficient functions, commonly denoted by expressions 
like ${\cal C}^{\rm S,(2)}_{i,g}(Q^2/\mu^2)$
(available in \cite{zn}, \cite{rijk}), with four-flavor light parton densities,
which also include a charm quark density $f_c(x,\mu^2)$. 
Since the corrections to the OME's contain terms in $\ln(Q^2/m^2)$ 
and $\ln(m^2/\mu^2)$ as well as nonlogarithmic terms it is simplest 
to work in the $\overline{\rm MS}$ scheme with the scale 
$\mu^2 = m^2$ for $Q^2 \le m^2$
and $\mu^2 = m^2 + Q^2(1-m^2/Q^2)^2/2$ for $Q^2 > m^2$ and discontinuous
matching conditions on the flavor densities at $\mu^2 = m^2$. 
Then all the logarithmic terms vanish at $Q^2=\mu^2=m^2$ 
and the non-logarithmic terms 
in the OME's are absorbed into the boundary conditions on the charm density,
the new four-flavor gluon density and the new light-flavor
u,d,s densities. The latter are convolutions of the previous 
three-flavor densities with the OME's given 
in the Appendix of \cite{bmsn1}. Hence we have a precise description 
through order $\alpha_s^2$ of how, in the limit $m\rightarrow 0$, 
to reexpress the $F_{i,c}^{\rm EXACT}(x,Q^2,m^2)$
written in terms of convolutions of heavy quark coefficient functions 
with three-flavor light parton densities into a description in terms of 
four-flavor light-mass parton coefficient functions
convoluted with four-flavor parton densities. 
This procedure leads to the so-called zero mass variable flavor number 
scheme (ZM-VFNS) for $F_{i,c}(x,Q^2,\Delta)$ where the $m$ dependent
logarithms are absorbed into the new four-flavor densities. 
To implement this scheme one has to be careful to use inclusive quantities 
which are collinearly finite in the limit $m \rightarrow 0$ and $\Delta$
is an appropriate parameter which enables us to do this. In the 
expression for $F_{i,c}$ there is a cancellation of terms 
in $\ln^3(Q^2/m^2)$ between the two-loop corrections to the light 
quark vertex function (the Sudakov form factor) and the convolution of 
the densities with the soft part of the $L_{2,q}$-coefficient function. 
This is the reason for the split of $L_{i,q}$ into soft and hard 
parts, via the introduction of a constant $\Delta$. Details
and analytic results for $L_{i,c}^{\rm SOFT}$ and $L_{i,c}^{\rm HARD}$ 
are available in \cite{csn1}.  All this analysis yielded 
and used the two-loop matching
conditions on variable-flavor parton densities across flavor thresholds,
which are special scales where one makes transitions from say a three-flavor
massless parton scheme to a four-flavor massless parton scheme. 
The threshold is a choice of $\mu$
which has nothing to do with the actual kinematical heavy flavour
pair production threshold at $Q^2(x^{-1} -1) = 4 m^2$. 
In \cite{bmsn1},\cite{bmsn2} it was shown that the
$F^{\rm EXACT}_{i,c}(x,Q^2,m^2, n_f=3)$ tend numerically to the 
known asymptotic results in $F^{\rm ASYMP}_{i,c}(x,Q^2,m^2,n_f=3)$,
when $Q^2 \gg m^2 $, which also equal the ZM-VFNS results, 
which were called $F^{\rm PDF}_{i,c}(x,Q^2,n_f=4)$.
The last description is good for large (asymptotic) scales and 
contains a charm density $f_c(x,\mu^2)$ which satisfies
a specific boundary condition at $\mu^2 = m^2$. 

\begin{figure}[h]
   \epsfig{file=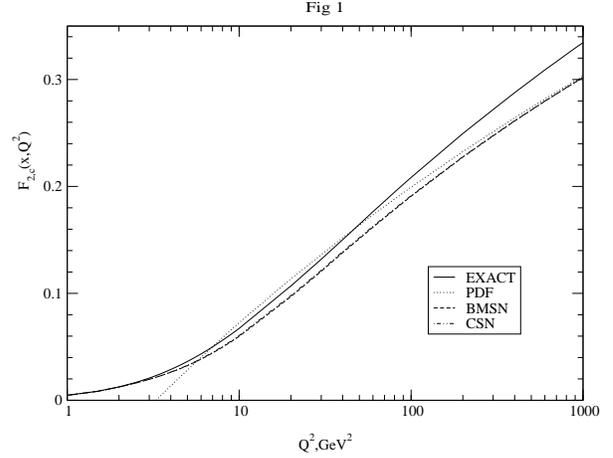,width=6cm,angle=270 }
\vspace{-0.8cm}
\caption{
The charm quark structure functions
$F_{2,c}^{\rm EXACT}(n_f=3)$ (solid line)
$F_{2,c}^{\rm CSN}(n_f=4)$, (dot-dashed line)
$F_{2,c}^{\rm BMSN}(n_f=4)$, (dashed line) and
$F_{2,c}^{\rm PDF}(n_f=4)$, (dotted line)
in NNLO for $x=0.005$
plotted as functions of $Q^2$.
}
\vspace{-0.4cm}
\end{figure}                                                              
s most of the experimental data occur in the kinematical regime 
between small $Q^2$ and asymptotic $Q^2$ a third approach has been 
introduced to describe the charm components of $F_{i}(x,Q^2)$.
This is called a variable flavor number scheme (VFNS). A first 
discussion was given in \cite{acot}, where a VFNS prescription 
called ACOT was given in lowest order only. A proof of factorization 
to all orders was recently given in \cite{col} for the total structure 
functions $F_{i}(x,Q^2)$, but the NLO expressions for $F_{i,c}(x,Q^2,m^2)$ 
in this scheme were not provided. An NLO version of a VFNS scheme has 
been introduced in \cite{csn1} and will be called the CSN scheme.
A different approach, also generalized to all orders, 
was given in \cite{bmsn1},\cite{bmsn2}, which is called the BMSN scheme. 
Finally another version of a VFNS was presented 
in \cite{thro}, which is called the TR scheme. 
The differences between the various schemes can be attributed to
two ingredients entering the construction of a VFNS. 
The first one is the mass factorization procedure carried out before the 
large logarithms can be resummed. The second one is the matching 
condition imposed on the charm quark density, which has to vanish 
in the threshold region of the production process. 
All VFNS approachs require two sets of parton densities. One set 
contains three-flavor number densities whereas the second set
contains four-flavor number densities. The sets have to satisfy the 
$\overline{\rm MS}$ matching relations derived in \cite{bmsn1}. 
Appropriate four-flavor densities have been constructed in \cite{csn1} 
starting from the three-flavor LO and NLO sets of parton densities recently 
published in \cite{grv98}.

\begin{figure}[h]
   \epsfig{file=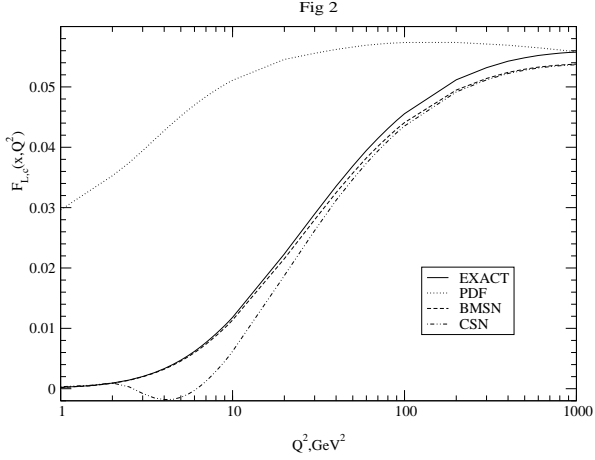,width=6cm,angle=270 }
\vspace{-0.8cm}
\caption{
The charm quark structure functions
$F_{L,c}^{\rm EXACT}(n_f=3)$ (solid line)
$F_{L,c}^{\rm CSN}(n_f=4)$, (dot-dashed line)
$F_{L,c}^{\rm BMSN}(n_f=4)$, (dashed line) and
$F_{L,c}^{\rm PDF}(n_f=4)$, (dotted line)
in NNLO for $x=0.005$
plotted as functions of $Q^2$.
}
\vspace{-0.4cm}
\end{figure}                                                              

Since the formulae for the heavy quark structure functions
are available in \cite{csn1} we only mention a few points here. 
The BMSN scheme avoids the introduction of any new coefficient
functions other than those above. 
Since the asymptotic limits for $Q^2 \gg m^2$ of all the operator matrix 
elements and coefficient functions are known we define (here $Q$ refers to the
heavy charm quark)
\begin{eqnarray}
&&F_{i,Q}^{\rm BMSN}(x,Q^2,m^2,\Delta,n_f=4)=
\nonumber\\[2ex]
&&
F_{i,Q}^{\rm EXACT}(x,Q^2,m^2,\Delta,n_f=3) 
\nonumber\\[2ex]
&& - F_{i,Q}^{\rm ASYMP}(x,Q^2,m^2,\Delta,n_f=3)
\nonumber\\[2ex]
&&
+ F_{i,Q}^{\rm PDF}(x,Q^2,m^2,\Delta,n_f=4)\,.
\end{eqnarray}
The scheme for $F_{i,Q}^{\rm CSN}$ introduces a new heavy quark OME 
$A_{QQ}^{\rm NS,(1)}(z,\mu^2/m^2)$ \cite{neve} and coefficient functions
$H_{i,Q}^{\rm NS,(1)}(z,Q^2/m^2)$ \cite{bnb}
because it requires an incoming heavy
quark $Q$, which did not appear in the NLO corrections in \cite{lrsn}.
The CSN coefficient fuctions are defined via the following equations. 
Up to second order we have 
\begin{eqnarray}
&& {\cal C}_{i,q,Q}^{\rm CSN,SOFT,NS,(2)}
\Big (\Delta,\frac{Q^2}{m^2},\frac{Q^2}{\mu^2}\Big) = 
\nonumber\\[2ex]
&&
A_{qq,Q}^{\rm NS,(2)}\Big(\frac{\mu^2}{m^2}\Big) 
{\cal C}_{i,q}^{\rm NS,(0)}
-\beta_{0,Q} \ln \left ( \frac{\mu^2}{m^2} \right ) 
\nonumber\\[2ex]
&& \times {\cal C}_{i,q}^{\rm NS,(1)} \Big(\frac{Q^2}{\mu^2}\Big)
-{\cal C}_{i,q}^{\rm VIRT,NS,(2)}(\frac{Q^2}{m^2})
\nonumber\\[2ex]
&& - L_{i,q}^{\rm SOFT,NS,(2)}\Big 
(\Delta,\frac{Q^2}{m^2},\frac{Q^2}{\mu^2}\Big) \,, 
\end{eqnarray}
with the virtual term the second order Sudakov form factor.
The other CSN coefficient functions are defined by 
equations like (we only give one of the longitudinal terms
for illustration)
\begin{eqnarray}
&&{\cal C}_{L,g}^{\rm CSN,S,(1)}\Big (\frac {Q^2}{m^2},\frac {Q^2}{\mu^2}\Big)
=
\nonumber\\[2ex]
&&
H_{L,g}^{\rm S,(1)}\Big (\frac {Q^2}{m^2}\Big)-
A_{Qg}^{\rm S,(1)}(\frac{\mu^2}{m^2})\,
{\cal C}_{L,Q}^{\rm CSN,NS,(0)}\Big (\frac {Q^2}{m^2}\Big) \,,
\end{eqnarray}
with 
${\cal C}_{L,Q}^{\rm CSN,NS,(0)}={4m^2}/{Q^2}$.
The CSN and BMSN schemes are designed to have the following two properties. 
First of all, suppressing unimportant labels, 
\begin{eqnarray}
&&F_{i,Q}^{\rm CSN}(n_f=4) = F_{i,Q}^{\rm BMSN}(n_f=4)
= 
\nonumber\\[2ex]
&&
F_{i,Q}^{\rm EXACT}(n_f=3) \quad \mbox{for}
\quad Q^2 \le m^2 \,.
\end{eqnarray}
Since $f_Q(m^2)^{\rm NNLO} \not = 0$ (see \cite{bmsn1}) this condition can be
only satisfied when we truncate the perturbation series at the same order.
The second requirement is that 
\begin{eqnarray}
&& \mathop{\mbox{lim}}\limits_{\vphantom{
\frac{A}{A}} Q^2 \gg m^2}
F_{i,Q}^{\rm BMSN}(n_f=4)
= 
\nonumber\\[2ex]
&&
\mathop{\mbox{lim}}\limits_{\vphantom{
\frac{A}{A}} Q^2 \gg m^2} F_{i,Q}^{\rm CSN}(n_f=4) 
= \mathop{\mbox{lim}}\limits_{\vphantom{\frac{A}{A}} Q^2 \gg m^2}
F_{i,Q}^{\rm PDF}(n_f=4) \,.
\end{eqnarray}
The only differences between the two schemes arises from 
terms in $m^2$ so they may not be equal 
just above $Q^2 = m^2$. This turns out to be the case for the
longitudinal structure function, which is more sensitive to mass effects.
We plot in Fig.1 NNLO results for the $Q^2$
dependence of $F_{2,c}^{\rm EXACT}(n_f=3)$,
$F_{2,c}^{\rm CSN}(n_f=4)$, $F_{2,c}^{\rm BMSN}(n_f=4)$, and 
$F_{2,c}^{\rm PDF}(n_f=4)$ at $x=0.005$. This figure shows that the results
satisfy the requirements in Eqs.(4) and (5). The ZM-VFNS description
is poor at small $Q^2$.
In Fig.2 we show the results for $F_{L,c}^{\rm EXACT}(n_f=3)$,
$F_{L,c}^{\rm CSN}(n_f=4)$, $F_{L,c}^{\rm BMSN}(n_f=4)$, and 
$F_{L,c}^{\rm PDF}(n_f=4)$ at $x=0.005$. We see that the CSN result
is negative and therefore unphysicsl 
for $2.5 < Q^2 < 6$ $({\rm GeV}/c)^2$ which is due
to the term in $4m^2/Q^2$ and the subtraction in Eq.(3). 

One way this research work is of relevance to Fermilab experiments 
is that it produces more precise ZM-VFNS parton densities.
Such densities are used extensively to predict cross sections 
at high energies, for example for 
single top quarks. Therefore the previous work on four-flavor parton 
densities has been extended in \cite{cs1}
to incorporate the two-loop discontinuous 
matching conditions across the bottom flavour threshold 
at $\mu = m_b$ and provided a set of five-flavour densities,
which contains a bottom quark density $f_b(x,\mu^2)$.
The differences between the five-flavor densities and those
in \cite{mrst98} and \cite{cteq5} are also discussed. 
Results for deep-inelastic electroproduction of bottom quarks 
are presented in \cite{csn2}.

\end{document}